\documentclass[dvips]{article}

\usepackage{icrctc07}

\title{Feasibility of acoustic neutrino detection in ice:
\\First results from the South Pole Acoustic Test Setup (SPATS)}

\shorttitle{First results from SPATS}

\authors{S.~B\"{o}ser$^1$, C.~Bohm$^2$, F.~Descamps$^3$, J.~Fischer$^1$, A.~Hallgren$^4$, R.~Heller$^1$, S.~Hundertmark$^2$, K.~Krieger$^1$, R.~Nahnhauer$^1$, M.~Pohl$^1$, P.~B.~Price$^5$, K.-H.~Sulanke$^1$, D.~Tosi$^1$, and J.~Vandenbroucke$^5$}

\shortauthors{J. Vandenbroucke and et al}

\afiliations{$^1$ DESY, Platanenallee 6, D-15738 Zeuthen, Germany\\
$^2$ University of Stockholm, Fysikum, SE-10691 Stockholm, Sweden\\
$^3$ University of Ghent, Department of Subatomic and Radiation Physics, Belgium\\
$^4$ University Uppsala, Department of Nuclear and Particle Physics, Box 535 SE-751 21 Uppsala, Sweden\\
$^5$ Dept. of Physics, University of California, Berkeley, CA 94720, USA}

\email{justinav@berkeley.edu}

\abstract{Astrophysical neutrinos in the EeV range (particularly those generated by the interaction of cosmic rays with the cosmic microwave background) promise to be a valuable tool to study astrophysics and particle physics at the highest energies.  Much could be learned from temporal, spectral, and angular distributions of $\sim$100 events, which could be collected by a detector with $\sim$100~km$^3$ effective volume in a few years.  Scaling the optical Cherenkov technique to this scale is prohibitive.  However, using the thick ice sheet available at the South Pole, the radio and acoustic techniques promise to provide sufficient sensitivity with sparse instrumentation. The best strategy may be a hybrid approach combining all three techniques.  A new array of acoustic transmitters and sensors, the South Pole Acoustic Test Setup, was installed in three IceCube holes in January 2007.  The purpose of SPATS is to measure the attenuation length, background noise, and sound speed for 10-100~kHz acoustic waves.  Favorable results would pave the way for a large hybrid array.  SPATS is the first array to study the possibility of acoustic neutrino detection in ice, the medium expected to be best for the purpose.  First results from SPATS are presented.}

\begin{document}
\maketitle

\begin{figure}
\begin{center}
\includegraphics [width=0.35\textwidth]{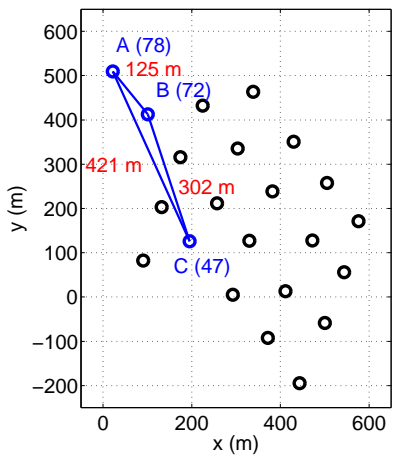}
\end{center}
\caption{SPATS layout (triangle).  All 22 IceCube strings installed as of July 2007 are shown.  SPATS (IceCube) string ID's are indicated.}\label{icrc1282_fig01}
\end{figure}

\section{Introduction}

An observatory capable of detecting $\sim$100 cosmogenic neutrinos in several years of operation promises to address fundamental questions of astrophysics and particle physics.  A possible realization of such a detector is to combine the optical, radio, and acoustic methods in a hybrid array in South Pole ice~\cite{Besson05}.  The optical method has detected thousands of neutrinos, and the radio method has been used to set the most stringent flux limits on the highest energy neutrinos. The acoustic method, on the other hand, is still in early development and the first step is to determine whether the acoustic properties of the ice are as favorable as has been predicted.

The South Pole Acoustic Test Setup (SPATS) was built to measure the attenuation length, ambient noise, and sound speed at South Pole.  The attenuation length is predicted to be several km by extrapolating a model that explains laboratory data~\cite{Price06}.  Ambient noise is expected to be quieter than in ocean water due to the lack of animal, human, and wave activity.  The sound speed was previously measured in the upper 186~m in the $\sim$Hz range~\cite{Weihaupt63}.

SPATS consists of three independent strings in IceCube \cite{Karle07} holes.  Each string has seven stages, with one each at 80, 100, 140, 190, 250, 320, and 400~m depth (Fig.~\ref{icrc1282_fig01}).  Each stage consists of a piezoelectric transmitter and a sensor module with 3 piezoelectric channels.  The SPATS array was deployed in January 2007 and has been taking data continuously since then.  For a description of the SPATS technical design and performance, see~\cite{Descamps07}.  In the same month three IceCube holes were also instrumented with radio receivers and transmitters as part of the AURA project~\cite{Landsman07}.  This includes the first hole instrumented with optical, radio, and acoustic sensors (IceCube Hole 78), a milestone toward a possible large hybrid array.

\begin{figure}
\begin{center}
\includegraphics [width=0.49\textwidth]{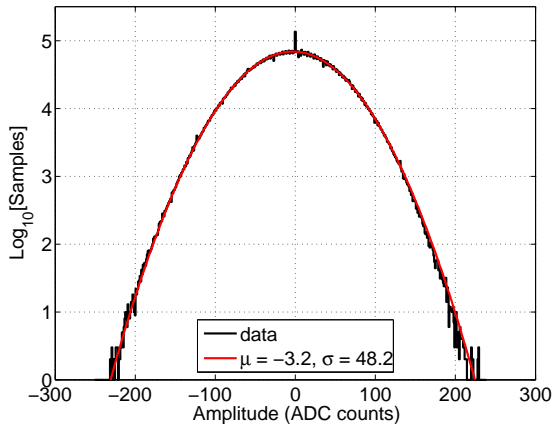}
\end{center}
\caption{ADC voltages on a single sensor channel, recorded intermittently over May 2007.  Histogram and Gaussian fit are shown.}\label{icrc1282_fig02}
\end{figure}

\section{Background noise}

The acoustic noise floor in the relevant bandwidth (10-100 kHz) determines the minimum amplitude pulse detectable, which for a given array design determines the neutrino energy threshold.  A low noise floor permits a low neutrino energy threshold, while a stable noise floor permits straightforward triggering and analysis.

Ocean and lake water, where all previous acoustic neutrino studies have been performed, suffers from a noise floor that is highly variable on a wide range of time scales~\cite{Vandenbroucke05}.  SPATS measurements, in contrast, indicate a noise floor that is Gaussian (Fig.~\ref{icrc1282_fig02}) and stable on all time scales from minutes to the entire observation period (5 months).

The amplitude $\sigma$ of the Gaussian noise is observed to decrease slightly with depth on each string.  This is consistent with the expectation that much of the ambient noise originates from human and wind activity on the surface and decreases with depth due to waveguiding in the upper $\sim$190~m.

The noise level at each channel increased slightly (by a few percent), and steadily, from February to April 2007, and then leveled off.  It has not varied in bursts as weather-induced surface noise would.  The strings were deployed by melting a column of ice and then allowing the water to refreeze surrounding the equipment.  This refrozen ice is known to asymptotically return to ambient ice temperature over several months, which may explain the initial noise increase: as the ice cooled it stiffened, better coupling the ambient noise from the bulk ice to the sensors.

\begin{figure}
\begin{center}
\includegraphics [width=0.49\textwidth]{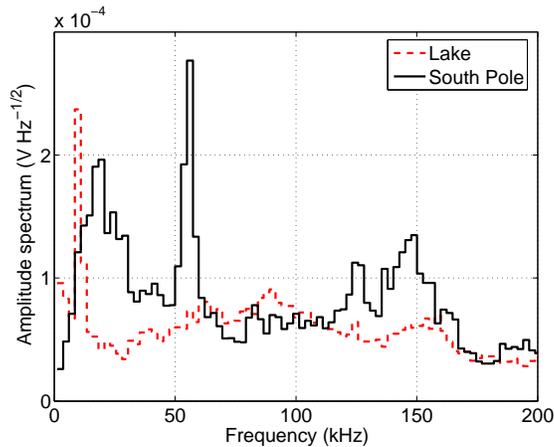}
\end{center}
\caption{Noise spectra for channel 0 of sensor module 21, in a lake and in South Pole ice.}\label{icrc1282_fig03}
\end{figure}

Example noise spectra are shown in Fig.~\ref{icrc1282_fig03}.  The sensors were tested at a Swedish lake, when the lake was covered with 90 cm of ice.  Three new features are present after installation in South Pole ice: a bump at $\sim$20~kHz, a bump at $\sim$150~kHz, and a spike at $\sim$55~kHz.  The features could be acoustic noise due to cracking ice during IceCube hole refreeze and/or glacial ice sheet movement.  The spectral features (as well as the Gaussian amplitude) are unaffected by powering off AMANDA, IceCube, and IceTop, but could be due to electromagnetic interference from another experiment or South Pole station electronics.  On each string the spike amplitude increases with depth.  The frequency at which the spike occurs is independent of sensor module on each string but varies from string to string.

By running with a discriminator trigger (threshold $\sim$5~$\sigma$), transient background noises are recorded above the Gaussian background.  Roughly one event per channel per minute is triggered.

\section{Transmitter runs}

The SPATS transmitters can be pulsed at up to 10's of Hz and recorded with sensors on the same or different strings.  Running in water immediately after installation of each string, each transmitter could be heard with multiple sensors on the same string.  After freeze-in, only the sensor on the same stage (30 cm below the transmitter) can hear it.  This is likely because the cable, other instrumentation, and the emission pattern of the transmitter prevent direct transmission to other stages on the same string, while the water/ice boundary (with an impedance ratio of $\sim$2.3) permitted multiple reflections guiding the signal to the other stages.

\begin{figure}
\begin{center}
\includegraphics [width=0.49\textwidth]{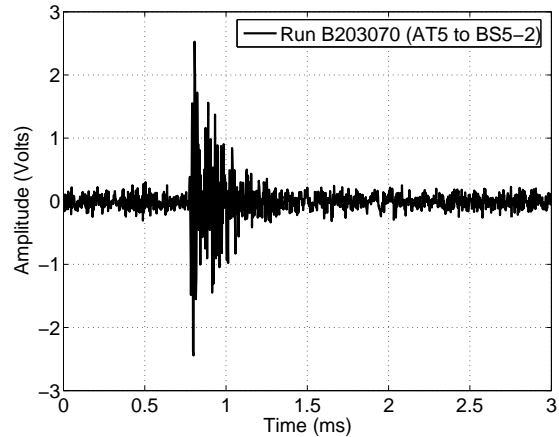}
\end{center}
\caption{A transmitter pulse produced on String A and recorded on String B.}\label{icrc1282_fig04}
\end{figure}

Between strings, transmitter pulses have been detected from each string to every other string (Fig.~\ref{icrc1282_fig04}).  However, in one-second runs with 18 pulses each, we have detected only 170 of the possible 882 inter-string transmitter-sensor pairs, and the number of pairs detected decreases with transmitter-sensor distance.  It was expected that the signal amplitude would decrease with distance due to $1/r$ divergence, but that all combinations would be detectable.  A signal to noise ratio of $\sim$10 was measured at 800~m in the lake, while the signal to noise ratio reaches 1 at $\sim$500~m in ice.  Accounting for the different noise levels, this means the recorded transmitter signals are an order of magnitude lower at South Pole than in the lake test.

This difference could be due to a difference in transmitter/sensor performance or coupling, or to a difference in attenuation during propagation (the lake attenuation length is likely several km).   Calculations and finite element simulations indicate that low temperature and high pressure improve the performance of both transmitters and sensors.  However, it is difficult to calibrate the transmitter and sensor response in their deployed conditions.  The module response was calibrated in water, and basic functionality was tested at low temperature.  But the response was not calibrated in ice, at simultaneous high pressure and low temperature.

The ``hole ice'' (column of refrozen ice surrounding each string) is known to have poor optical properties due to its high concentration of air bubbles.  Its acoustic properties are unknown.  Bubbles on transmitter and sensor elements could have a significant effect.  A small ($\sim$100~$\mu$m or larger) air gap between a sensor module and the surrounding ice is expected to severely attenuate signals.  This is because the signal cannot penetrate the air layer, which has an impedance $\sim10^{-4}$ that of ice.  There could also be cracks in or surrounding the hole column from the freeze-in process.  If so, they should anneal slowly over time.

\begin{figure}
\begin{center}
\includegraphics [width=0.49\textwidth]{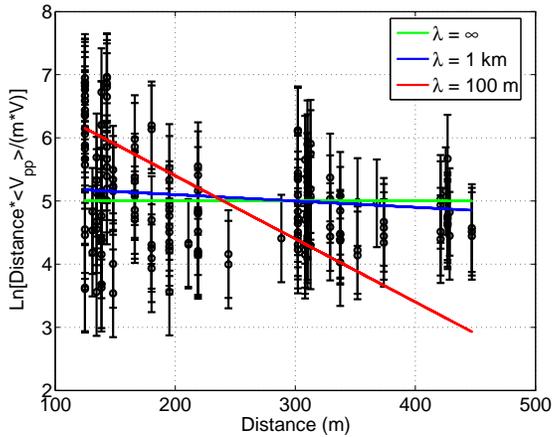}
\end{center}
\caption{Amplitude multiplied by distance (to compensate $1/r$ dependence) vs. distance, for every transmitter - sensor channel combination with a transmitter pulse detected above noise.  Model curves are shown for different attenuation lengths.}\label{icrc1282_fig05}
\end{figure}

The factor of $\sim$10 change in module performance/coupling appears to be an overall normalization effect and not a bulk-ice attenuation effect.  Fig.~\ref{icrc1282_fig05} shows amplitude times distance vs. distance for the 170 detected inter-string transmitter-sensor combinations.  Each amplitude was determined by recording a sensor channel continuously with a transmitter pulsing, averaging 18 recorded pulses, and then determining the peak-to-peak voltage.  Error bars show estimated systematic error, which is dominated by module-to-module variation in coupling.  There is large scatter in the data, which is due both to noise and to module-to-module variation.  The current data set does not significantly constrain the attenuation length.

\section{Conclusion}

The South Pole Acoustic Test Setup was successfully installed and commissioned between December 2006 and February 2007.  All 21 transmitters and 53 of 63 sensor channels are working normally.  Data taking is ongoing.  Ambient noise is stable and Gaussian, and decreases with depth.  It increased slightly in the first 3 months and then leveled off.  Transmitter signals have been detected from each string to every other string.  Their amplitude is an order of magnitude lower than expected.  The current data do not significantly constrain the attenuation length.  Data will be taken next with 100 times as many transmitter pulses, averaging over which will improve the signal to noise ratio by a factor of 10.  This should allow us to detect new combinations to constrain the attenuation length.  The timing of the transmitter pulses will also be analyzed both to measure the sound speed and to determine the effect of refraction on the attenuation analysis.

\section{Acknowledgements}
We are grateful for the support of the IceCube collaboration.  J.~Vandenbroucke acknowledges support from the National Science Foundation GRFP.

%This is the reference to .bib file (Whitout .bib!)
\bibliography{icrc1282}
%This in the bibtex style, is ok.
\bibliographystyle{unsrt}

\end{document}